\documentclass[journal, 10pt]{IEEEtran}
\newcommand{\rev}[1]{{{#1}}}
\usepackage{color, soul}

\usepackage{subcaption}

\usepackage{amsfonts}
\usepackage{amsmath}
\usepackage{amssymb}
\usepackage[linesnumbered,ruled,vlined]{algorithm2e}
\SetKwRepeat{Do}{do}{while}%
\usepackage{amsthm}
\usepackage{algpseudocode}

\DeclareMathOperator{\Unif}{Unif}

\usepackage{graphicx}
\usepackage{hyperref}

\begin{document}
\title{An approximate expectation-maximization for two-dimensional multi-target detection}
\author{Shay Kreymer, Amit Singer, and Tamir Bendory
\thanks{S. Kreymer and T. Bendory are with the School of Electrical Engineering of Tel Aviv University, Tel Aviv, Israel, e-mail: \href{mailto:shaykreymer@mail.tau.ac.il}{shaykreymer@mail.tau.ac.il}, \href{mailto:bendory@tauex.tau.ac.il}{bendory@tauex.tau.ac.il}. A. Singer is with the Department of Mathematics and PACM, Princeton University, Princeton, NJ, USA, e-mail: \href{mailto:amits@math.princeton.edu}{amits@math.princeton.edu}. S.K. is supported by the Yitzhak and Chaya Weinstein Research Institute for Signal Processing. A.S. is partly supported by AFOSR Award FA9550-20-1-0266, the Simons Foundation Math+X Investigator Award, NSF BIGDATA Award IIS1837992, NSF Award DMS-2009753, and NIH/NIGMS Award R01GM136780- 01. T.B. is partially supported by the BSF grant no. 2020159, the NSF-BSF grant no. 2019752, and the ISF grant no. 1924/21.}}

\markboth{IEEE Signal Processing Letters}%
{Kreymer \MakeLowercase{\textit{et al.}}: An approximate expectation-maximization for two-dimensional multi-target detection}

\maketitle
\begin{abstract}
We consider the two-dimensional multi-target detection (MTD) problem of estimating a target image from a noisy measurement that contains multiple copies of the image, each randomly rotated and translated. The MTD model serves as a mathematical abstraction of the structure reconstruction problem in single-particle cryo-electron microscopy, the chief motivation of this study. We focus on high noise regimes, where accurate detection of image occurrences within a measurement is impossible. To estimate the image, we develop an  expectation-maximization framework that aims to maximize an approximation of the  likelihood function. We demonstrate image recovery in highly noisy environments, and show that our framework outperforms the previously studied autocorrelation analysis in a wide range of parameters.
\end{abstract}

\setlength{\textfloatsep}{5pt}

\section{Introduction}
\label{sec:introduction}
We study the multi-target detection (MTD) problem of estimating a target image~\mbox{$f:\mathbb{R}^2 \rightarrow \mathbb{R}$} from a noisy measurement that contains multiple copies of the image, each randomly rotated and translated~\cite{bendory2019multi, lan2020multi, marshall2020image, bendory2021multi, kreymer2021two, shalit2021generalized, bendory2018toward}. We consider a measurement~$M \in \mathbb{R}^{N \times N}$ of the form
\begin{equation}
\label{eq:model}
M[\vec{\ell}] = \sum_{i=1}^{p} F_{\phi_i}[\vec{\ell} - \vec{\ell}_i] + \varepsilon[\vec{\ell}],
\end{equation}
where \mbox{$F_{\phi_i} [\vec{\ell}] := f_{\phi_i} (\vec{\ell} / n)$} is a discrete copy of~$f$, rotated by angle~$\phi_i$ about the origin;~$n$ is the \rev{known} radius of the image in pixels; \mbox{$\{\phi_i\}_{i=1}^{p} \sim \Unif[0, 2\pi)$} are uniformly distributed rotations; \mbox{$\{\vec{\ell}_i\}_{i=1}^{p} \in \{n + 1, \ldots, N-n\}^2$} are \rev{identically distributed random translations (though not independent, and the underlying distribution is not assumed to be known)}; and $\varepsilon[\vec{\ell}]$ is i.i.d.\ Gaussian noise with zero mean and \mbox{variance~$\sigma^2$}. The rotations, translations, and the number of occurrences of~$f$ in~$M$, denoted by~$p$, are unknown. Importantly, since the rotations are unknown, it is possible to reconstruct the target image only up to a rotation.

Following~\cite{marshall2020image,bendory2021multi,kreymer2021two,zhao2013fourier}, we assume that the image~$f$ is supported on the unit disk \mbox{$\{\vec{x} \in \mathbb{R}^2: |\vec{x}| \le 1\}$}
and has a finite expansion in the basis of Dirichlet Laplacian eigenfunctions. In particular, the image~$f$ can be expanded as
\begin{equation}
\label{eq:expansion}
f (r, \theta) = \sum_{(\nu, q): \lambda_{\nu, q} \le \lambda} \alpha_{\nu, q} \psi_{\nu, q} (r, \theta), \quad r \le 1,
\end{equation}
in polar coordinates $(r, \theta)$, where~$\psi_{\nu, q}(r,\theta) = J_\nu\left( \lambda_{\nu, q} r \right) e^{i \nu \theta}$, \mbox{$\nu \in \mathbb{Z}_{\ge 0}$},~$J_\nu$ is the~\mbox{$\nu$-th} order Bessel function of the first kind, \mbox{$\lambda_{\nu, q} > 0$} is the~\mbox{$q$-th} positive root of~$J_\nu$,~$\lambda$ is called the bandlimit frequency, and~$\alpha$ is the vector of expansion coefficients. Hereafter, by estimating the image we mean estimating the vector of coefficients~$\alpha$. Notably, the basis of Dirichlet Laplacian eigenfunctions is steerable: rotating~$f$ is equivalent to modulating the expansion coefficients~$\alpha_{\nu, q}$. Specifically, the expansion of the rotated image~$f_\phi (r, \theta) := f(r, \theta + \phi)$ is given by
$f_\phi (r, \theta) = \sum_{(\nu, q): \lambda_{\nu, q} \le \lambda} \alpha_{\nu, q} \psi_{\nu, q} (r, \theta) e^{i \nu \phi}$. \rev{For real valued images,~$\alpha_{-\nu,q} = \alpha_{\nu,q}^*$.}

\begin{figure}[!tb]
	\begin{subfigure}[ht]{0.30\columnwidth}
		\centering
		\includegraphics[width=\columnwidth]{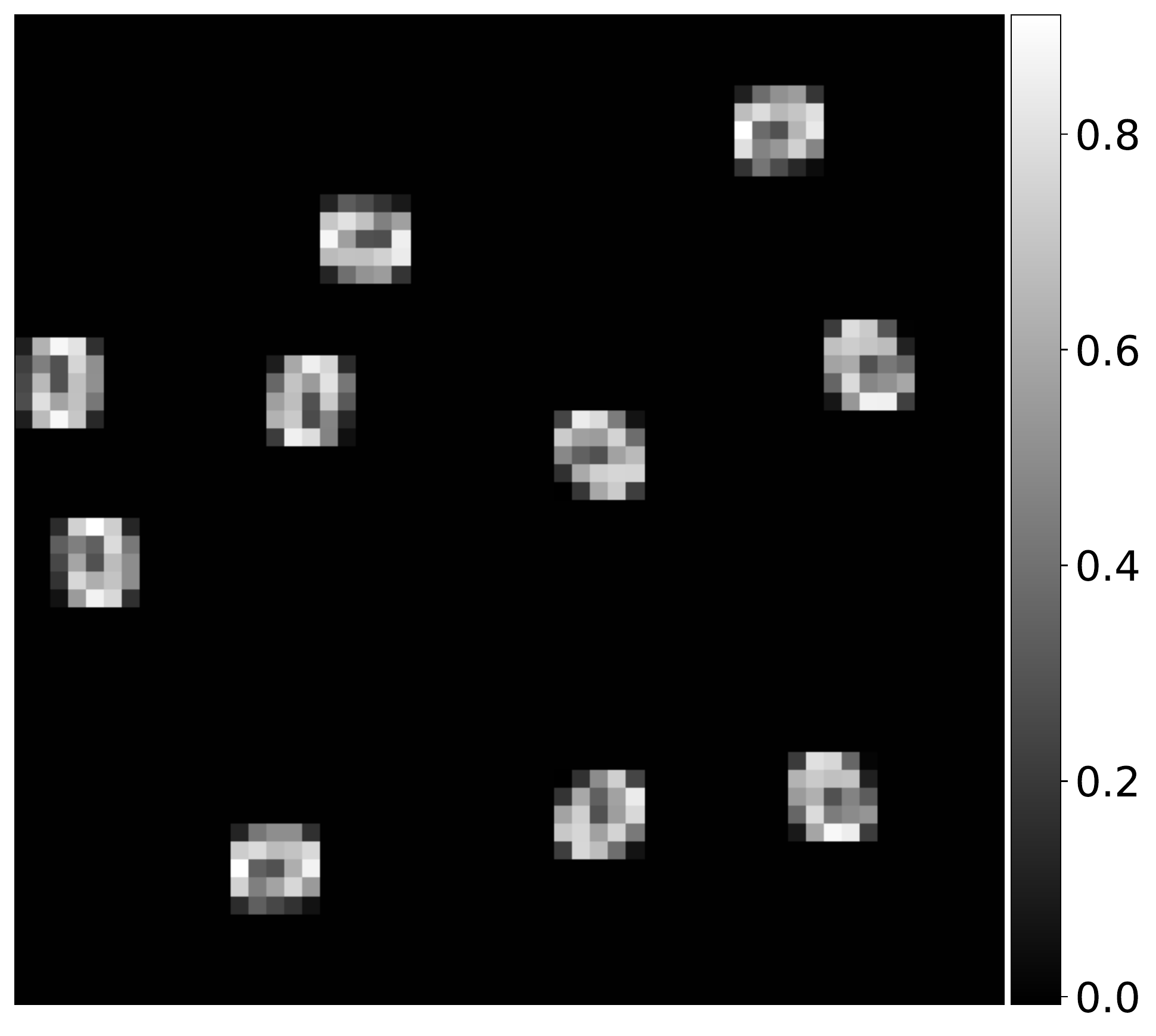}
		\caption{No noise}
	\end{subfigure}
	\hfill
	\begin{subfigure}[ht]{0.30\columnwidth}
		\centering
		\includegraphics[width=\columnwidth]{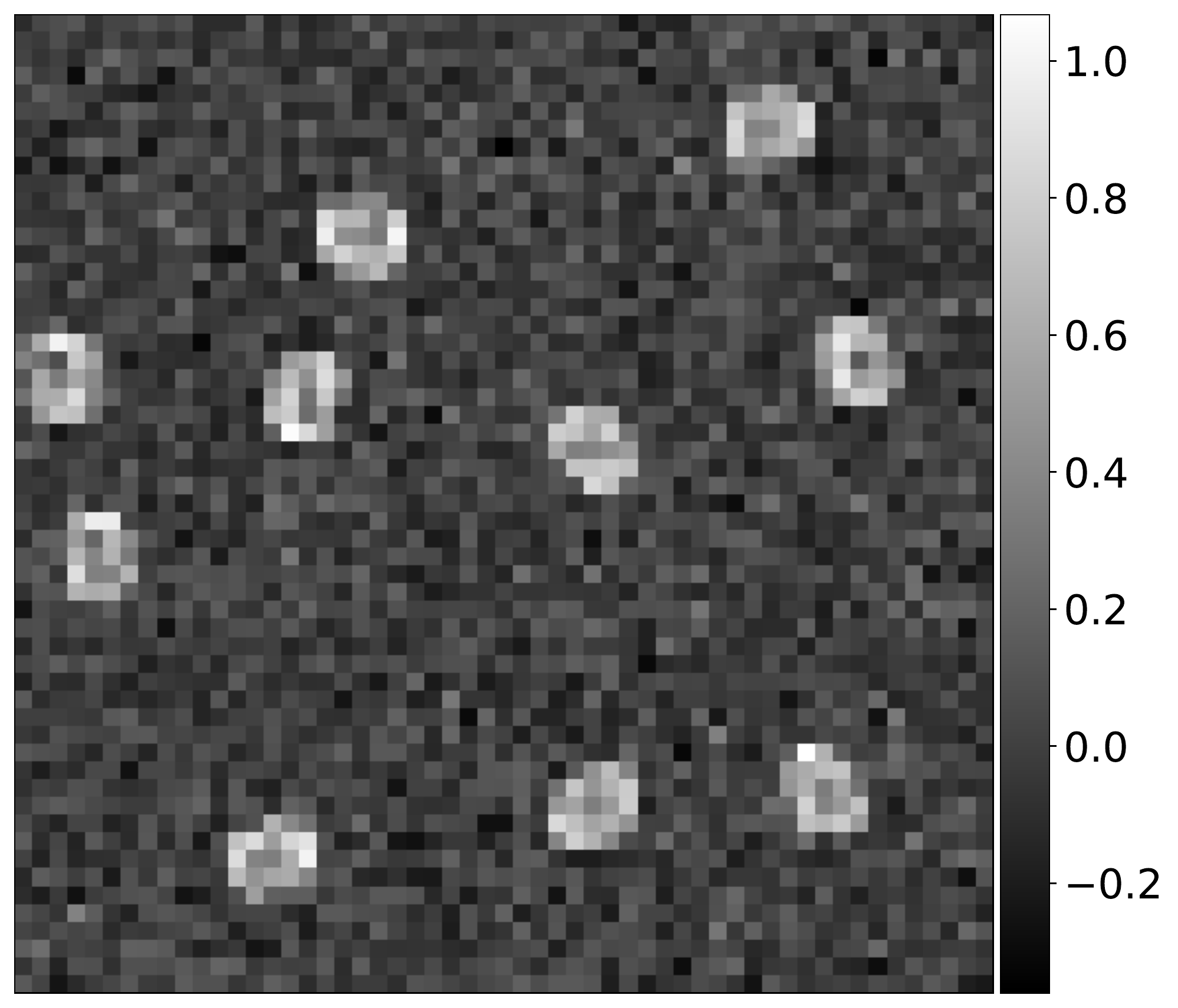}
		\caption{$\text{SNR} = 50$}
	\end{subfigure}
	\hfill
	\begin{subfigure}[ht]{0.30\columnwidth}
		\centering
		\includegraphics[width=\columnwidth]{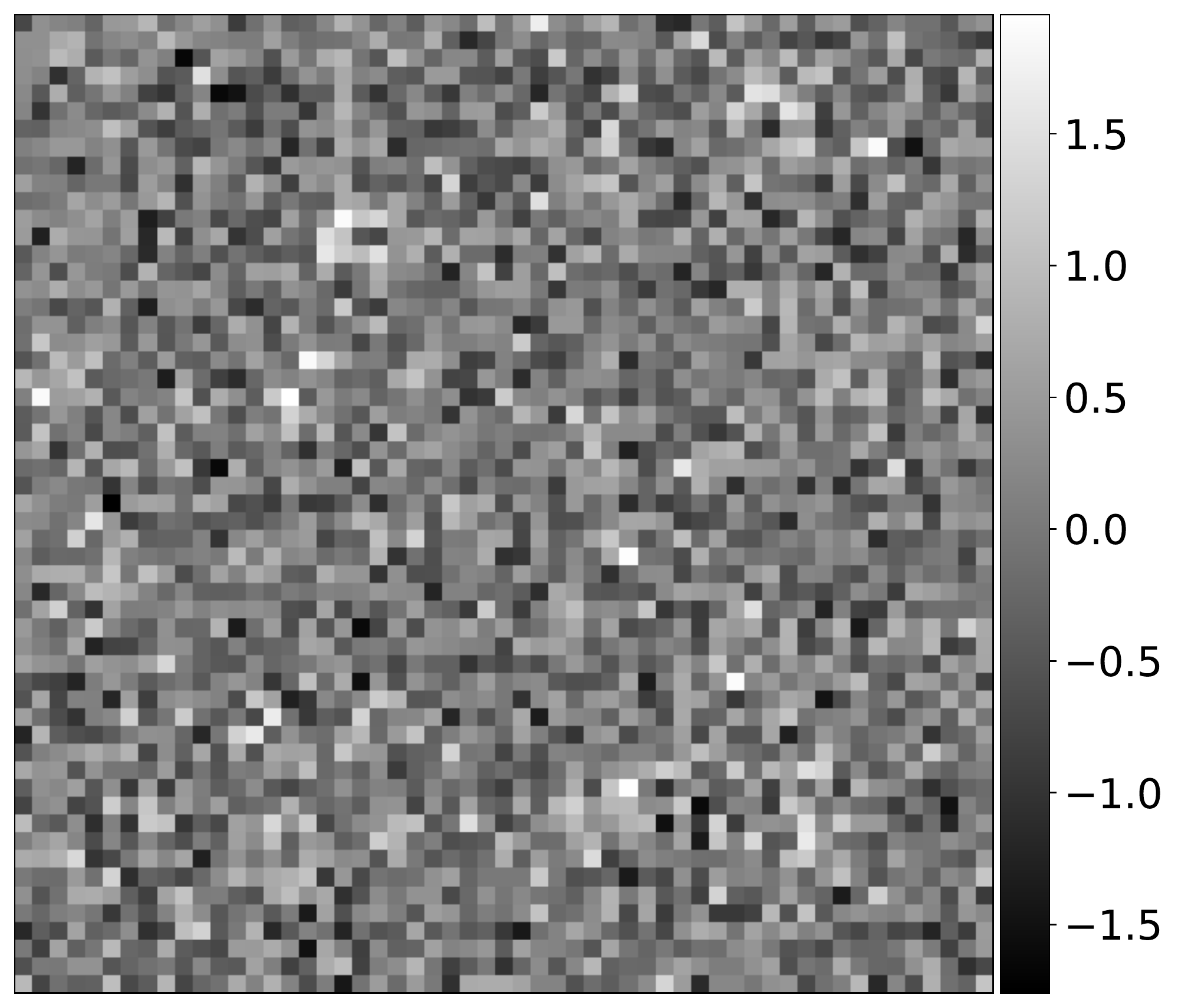}
		\caption{$\text{SNR} = 2$}
	\end{subfigure}
	\caption{Three measurements at different SNRs: (a)~no noise; (b)~\mbox{$\text{SNR} = 50$}; (c)~\mbox{$\text{SNR} = 2$}. Each measurement contains multiple rotated versions of the target image. We focus on the low SNR regime~{(e.g., panel~(c))} in which the locations and rotations of the image occurrences cannot be detected reliably.}
\label{fig:Micrographs_noise}
\end{figure}

We focus on the \mbox{well-separated} case of the 2-D MTD problem, which was introduced in~\cite{marshall2020image, bendory2021multi}. In this case, each translation in the measurement~$M$~\eqref{eq:model} is separated by at least a full image diameter from its neighbors. Specifically, we assume
\begin{equation}
\label{eq:sep}
|\vec{\ell}_{i_1} - \vec{\ell}_{i_2}| > 4n, \quad \text{ for all } i_1 \ne i_2.
\end{equation}
Figure~\ref{fig:Micrographs_noise} presents an example of a measurement~$M$ at different signal-to-noise ratios (SNRs). We define~\mbox{$\text{SNR} := \frac{\|F\|_\text{F}^2}{A \sigma^2}$}, where~$A$ is the area in pixels of~$F$.

The MTD model serves as a mathematical abstraction of the cryo-electron microscopy~(\mbox{cryo-EM}) technology for macromolecular structure determination~\cite{henderson1995potential, nogales2016development, bai2015cryo}. In a \mbox{cryo-EM} experiment, individual copies of the target biomolecule are dispersed at unknown \mbox{2-D} locations and \mbox{3-D} orientations in a thin layer of vitreous ice, from which \mbox{2-D} tomographic projection images are produced by an electron microscope~\cite{frank2006three}. It is necessary to keep the electron dose low in order to minimize  irreversible structural damage. Consequently, the projection images are considerably noisy. In the current data processing pipeline of \mbox{cryo-EM}~\cite{bendory2020single, singer2020computational, scheres2012relion, punjani2017cryosparc}, the~\mbox{2-D} projections are first detected and extracted from the micrograph, and later rotationally and translationally aligned to reconstruct the~\mbox{3-D} molecular structure. This approach fails for small molecules, which are difficult to detect and align~\cite{bendory2018toward, henderson1995potential, bendory2020single, aguerrebere2016fundamental}.

The MTD model was devised in \cite{bendory2018toward} in order to study the recovery of small molecular structures using cryo-EM, below the current detection limit~\cite{d2021current}. In~\cite{marshall2020image, bendory2021multi, kreymer2021two}, an autocorrelation analysis technique was devised for the 2-D MTD problem~\eqref{eq:model}. Autocorrelation analysis is a special case of the method of moments, and it consists of finding an image that best matches the empirical autocorrelations of the measurement, thus bypassing detecting the locations and rotations of individual image occurrences. In this work, we propose to replace autocorrelation analysis by expectation-maximization~(EM): a classical iterative algorithm to compute the  maximum likelihood estimator~\cite{dempster1977maximum}. Similarly to autocorrelation analysis, the EM algorithm estimates the target image~$F$ directly by marginalizing over the translations and rotations.

Previous works demonstrated that EM outperforms autocorrelation analysis in terms of estimation accuracy for the \mbox{1-D} MTD problem~\cite{lan2020multi}, as well as for the closely related multireference alignment model~\cite{bendory2017bispectrum, abbe2018multireference}. EM is also the most popular computational framework for reconstructing molecular structures using cryo-EM, see for example~\cite{scheres2012relion, punjani2017cryosparc}. Moreover, a recent paper~\cite{katsevich2020likelihood} shows that likelihood optimization in the low SNR regime reduces to a sequence of least squares optimization problems that match the moments of the image estimate to the observable moments one by one, and by that suggests that EM has the potential to surpass the estimation accuracy achieved by autocorrelation analysis.

At each iteration, EM assigns probabilities to all possible rotations and translations (see Section~\ref{sec:EM}). Unfortunately, for the MTD model~\eqref{eq:model}, the number of possible translations grows quickly with the measurement size, and therefore direct application of the EM algorithm to the MTD problem is computationally intractable, even for very small measurements. Thus, following~\cite{lan2020multi}, we suggest mitigating the computational burden by developing an EM algorithm that maximizes an approximation of the likelihood function. In the approximate EM scheme, the number of possible translations is linear in~$N^2$ (the size of the measurement), making the algorithm tractable.

The main contribution of this paper is in developing an approximate EM framework for the 2-D MTD problem; see Section~\ref{sec:EM}. In Section~\ref{sec:numeric}, we demonstrate successful reconstructions in noisy regimes. We also conduct extensive numerical experiments that  demonstrate significant improvement in estimation accuracy compared to autocorrelation analysis. Section~\ref{sec:conclusion} concludes the paper and introduces  future work. In particular, we discuss potential implications for the ongoing effort to estimate small molecular structures using cryo-EM~\cite{bendory2018toward}.

\section{Approximate expectation-maximization}
\label{sec:EM}
Given a measurement~$M$ that follows the MTD model~(\ref{eq:model}), the maximum marginal likelihood estimator~(MMLE) for the vector of coefficients~$\alpha$, that represents the target image~$f$~(\ref{eq:expansion}), is the maximizer of the likelihood function~$p(M | \alpha)$. The translations and rotations of the target images within the measurement are treated as nuisance variables. The EM algorithm estimates the MMLE by iteratively applying the expectation~(E) and maximization~(M) steps~\cite{dempster1977maximum}. Specifically, given the current estimate~$\alpha_k$, the \mbox{E-step} computes the expected log-likelihood function, where the expectation is taken over all admissible configurations of translations and rotations. The estimate is then updated in the \mbox{M-step} by maximizing the function with respect to~$\alpha$. Unfortunately, for the MTD model, the number of possible translations grows quickly with the problem size, rendering direct application of EM computationally intractable. Hence, based on~\cite{lan2020multi}, we suggest to apply an \rev{approximate EM}, in which the number of possible translations is linear in~$N^2$.

The approximate EM begins \rev{with} partitioning the measurement~$M$ into~$N_d = (N / L)^2$ \mbox{non-overlapping} patches; each patch is of size~\mbox{$L \times L$}, where~\mbox{$L = 2n + 1$} is the diameter of the target image~$F$. \rev{In EM terminology, the patches are called the observed data.} The separation condition~(\ref{eq:sep}) implies that each patch $M_m$ can contain either no target image, a full  target image, or part of a rotated image; overall there are~$(2L)^2$  possibilities\rev{, excluding rotations}. In particular, each patch can be modeled by
\begin{equation}
\label{eq:patch}
M_m = C T_{\vec{\ell}_m} Z F^{\rev{L}}_{\phi_m} + \varepsilon_m, \quad \varepsilon_m \sim \mathcal{N}(0, \sigma^2 I_{L \times L}),
\end{equation}
where~\rev{$F^L$ is a square image of size $L \times L$ enclosing the disk-shaped image~$F$ of radius~$n$}, the operator~$Z$ \mbox{zero-pads}~$L$ entries to the right and to the bottom of a rotated copy of~$F$, and~$T_{\vec{\ell}_m}$ circularly shifts the \mbox{zero-padded} image by~\mbox{$\vec{\ell}_m = ({\ell_m}_x, {\ell_m}_y)\in \mathbb{L} := \{0, 1, \ldots, 2L-1\}^2$} positions, that is,
\begin{multline}
(T_{\vec{\ell}_m} Z F^{\rev{L}}_{\phi_m} )\left[i, j\right] = \\(Z F^{\rev{L}}_{\phi_m}) \left[(i + {\ell_m}_x) \bmod 2L, (j + {\ell_m}_y) \bmod  2L\right].
\end{multline}
The operator~$C$ then crops \rev{the image to size $L \times L$, i.e., \mbox{$CT_{\vec{\ell}_m} Z F^{\rev{L}}_{\phi_m} = T_{\vec{\ell}_m} Z F^{\rev{L}}_{\phi_m}[0:L-1, 0:L-1]$}}, and the result is further corrupted by additive white Gaussian noise. In addition, since the EM algorithm assigns probabilities to rotations (in the expectation step), we need to discretize the search space of rotations:
\begin{equation}
\label{eq:Phi_set}
\phi_m \in \Phi := \left\{k \frac{2\pi}{K}\right\}, \quad k=0,\ldots,K-1,
\end{equation}
where~$K$ is a parameter chosen by the user. Higher~$K$ provides higher accuracy at the cost of running time (see Figure~\ref{fig:discretization_experiment}). \rev{The rotations, translations, and patches are referred to, in EM terminology, as the complete data.} We assume that in each patch~$p(\vec{\ell}_m,\phi_m|\alpha_k)=p(\vec{\ell}_m)p(\phi_m)$, namely,~$\vec{\ell}_m$ and~$\phi_m$ are independent of~$\alpha_k$ and of each other. Since the rotations are drawn from a uniform distribution over the set~$\Phi$ from~(\ref{eq:Phi_set}), we can write~$p(\vec{\ell}_m, \phi_m|\alpha_k) = \rho[\vec{\ell}_m]/K $, where~$\rho[\vec{\ell}]$ is the distribution of \mbox{2-D} translations in the patch (which should be estimated simultaneously with $\alpha$).

In the E-step, our algorithm calculates the expected log-likelihood function of the model, given explicitly by
\begin{equation}
Q(\alpha, \rho| \alpha_k, \rho_k) = \mathbb{E}\left[\log \mathfrak{L}| M_0, \ldots, M_{N_d-1}; \alpha_k, \rho_k \right],
\end{equation}
where $\mathfrak{L}$ is the likelihood function, defined as
\begin{multline}
\label{eq:approx_likelihood}
\mathfrak{L} := p(M_0, \ldots, M_{N_d-1}, \vec{\ell}_0, \ldots, \vec{\ell}_{N_d-1}, \phi_0, \ldots, \phi_{N_d-1};\alpha, \rho) \\\approx \prod_{m = 0}^{N_d - 1} p(M_m, \vec{\ell}_m, \phi_m;\alpha, \rho),
\end{multline}
where we neglect statistical dependencies between patches.

Bayes' rule dictates
\begin{multline}
p(\vec{\ell}_m, \phi_m|M_m, \alpha_k) \\= \frac{p(M_m|\vec{\ell}_m, \phi_m, \alpha_k) p(\vec{\ell}_m, \phi_m|\alpha_k)}{\sum_{\vec{\ell'} \in \mathbb{L}} \sum_{\phi' \in \Phi} p(M_m|\vec{\ell'}, \phi', \alpha_k) p(\vec{\ell'}, \phi'|\alpha_k)},
\end{multline}
which is the normalized likelihood function
\begin{equation}
\label{eq:likelihood_patch}
p(M_m|\vec{\ell}_m, \phi_m, \alpha) \propto \exp \left(- \frac{\|M_m - C T_{\vec{\rev{\ell}_m}} Z F_{\phi_m}\|_\text{F}^2}{2 \sigma^2} \right),
\end{equation}
with the normalization~\mbox{$\sum_{\vec{\ell} \in \mathbb{L}} \sum_{\phi \in \Phi} p(M_m|\vec{\ell}_m, \phi_m, \alpha) = 1$}, weighted by the prior distribution $p(\vec{\ell}_m, \phi_m| \alpha_k) = \rho[\vec{\ell}_m] / K$.

Utilizing the approximation in~(\ref{eq:approx_likelihood}), we can write the expected log-likelihood function, up to a constant, as:
\begin{multline}
\label{eq:Q_function}
Q(\alpha, \rho|\alpha_k, \rho_k) = \sum_{m = 0}^{N_d - 1} \sum_{\vec{\ell} \in \mathbb{L}} \sum_{\phi \in \Phi} p(M_m|\vec{\ell}, \phi, \alpha_k) \rho_k[\vec{\ell}] \\ \times \left(\log p(M_m|\vec{\ell}, \phi, \alpha) + \log \rho[\vec{\ell}]\right).
\end{multline}

The~\mbox{M-step} updates the image estimate~$\alpha$ and~$\rho$ by maximizing~$Q(\alpha, \rho|\alpha_k, \rho_k)$ under the constraint that~$\rho$ lies on the simplex~$\Delta_{4L^2}$:
\begin{equation}
	\label{eq:M_step}
	\arg \max_{\alpha, \rho} Q(\alpha, \rho|\alpha_k, \rho_k) \quad \text{s.t.} \quad  \rho \in \Delta_{4L^2}.
\end{equation}
The constrained maximization of~(\ref{eq:M_step}) can be achieved with the unconstrained maximization of the Lagrangian
\begin{equation}
\mathcal{L}(\alpha, \rho, \eta) = Q(\alpha, \rho|\alpha_k, \rho_k) + \eta \left(1 - \sum_{\vec{\ell} \in \mathbb{L}} \rho[\vec{\ell}] \right),
\end{equation}
where~$\eta$ is the Lagrange multiplier. As we will see next, the constraint of~\eqref{eq:M_step} is automatically satisfied at the maximum of the Lagrangian.

Since~$Q(\alpha, \rho|\alpha_k, \rho_k)$ is additively separable for~$\alpha$ and~$\rho$, we maximize~$\mathcal{L}(\alpha, \rho, \eta)$ with respect to~$\alpha$ and~$\rho$ separately. At the maximum of~$\mathcal{L}(\alpha, \rho, \eta)$, we have
\begin{multline}
\label{eq:update_alpha}
0 = \frac{\partial \mathcal{L}}{\partial{(\alpha)_{\nu, q}}} =  \sum_{m = 0}^{N_d - 1} \sum_{\vec{\ell} \in \mathbb{L}} \sum_{\phi \in \Phi} p(M_m|\vec{\ell}, \phi, \alpha_k) \rho_k[\vec{\ell}] \\ \times \frac{\partial \log p(M_m|\vec{\ell}, \phi, \alpha)}{\partial{(\alpha)_{\nu, q}}},
\end{multline}
resulting in a set of linear equations which is solved to update~$\alpha$. In order to update~$\rho$, we maximize~$\mathcal{L}(\alpha, \rho, \eta)$ with
respect to~$\rho$:
\begin{equation}
0 = \frac{\partial \mathcal{L}}{\mathcal \rho[\vec{\ell}]} = \sum_{m = 0}^{N_d - 1} \sum_{\phi \in \Phi} p(M_m|\vec{\ell}, \phi, \alpha_k) \rho_k[\vec{\ell}] \frac{1}{\rho[\vec{\ell}]} - \eta,
\end{equation}
for~$\vec{\ell} \in \mathbb{L}$. We thus obtain the update rule for~$\rho$ as
\begin{equation}
\label{eq:update_rho}
\rho[\vec{\ell}] = \frac{1}{\eta} \sum_{m = 0}^{N_d - 1} \sum_{\phi \in \Phi} p(M_m|\vec{\ell}, \phi, \alpha_k) \rho_k[\vec{\ell}],
\end{equation}
and~$\eta = N_d$ from the normalization~\mbox{$\sum_{\vec{\ell} \in \mathbb{L}} \rho[\vec{\ell}] = 1$}. The approximate EM algorithm is summarized in Algorithm~\ref{alg:approx_EM}.

\begin{algorithm}[!tb]
  \caption{Approximate EM for \mbox{2-D} MTD}\label{alg:approx_EM}
\KwIn{measurement~$M$; noise variance~$\sigma^2$; initial guesses~$\alpha_0$ and~$\rho_0$; parameter~$K$~\eqref{eq:Phi_set}; stopping parameter~$\epsilon$}
\KwOut{an estimate of~$\alpha$ and~$\rho$}
\BlankLine
set $k \rightarrow 0$;\\
calculate~$Q_{0}$ according to~(\ref{eq:Q_function}) and set $Q_{-1} \rightarrow -\infty$;\\
  \While{$Q_k - Q_{k-1} > \epsilon$}{
    calculate~$p(M_m|\vec{\ell}, \phi, \alpha_k)$ according to~(\ref{eq:likelihood_patch});\\
update~$\alpha_{k+1}$ by solving~(\ref{eq:update_alpha});\\
update~$\rho_{k+1}$ according to~(\ref{eq:update_rho});\\
calculate~$Q_{k+1}$ according to~(\ref{eq:Q_function});\\
set $k \rightarrow k + 1$;
  }
\end{algorithm}

\begin{figure}[!tb]
	\begin{subfigure}[ht]{\columnwidth}
		\centering
		\includegraphics[width=0.75\columnwidth]{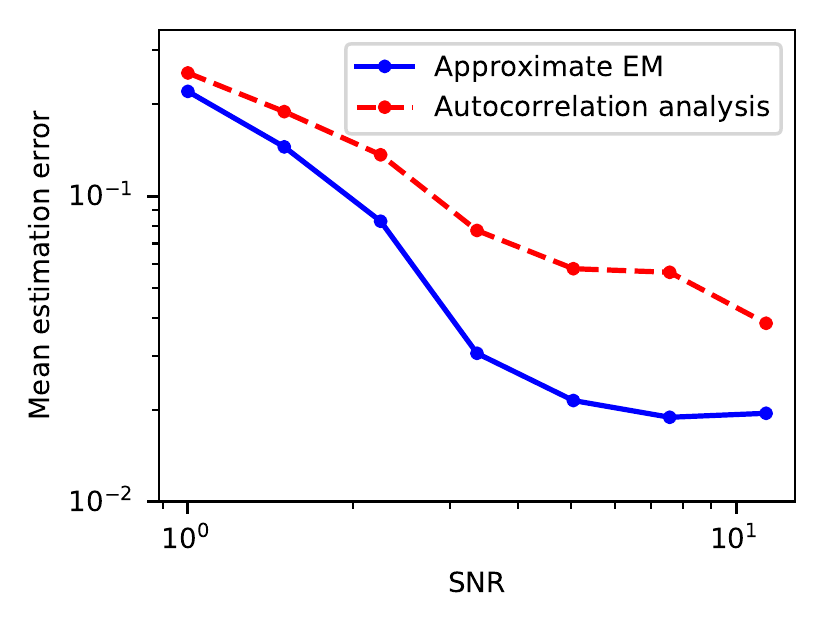}
	\end{subfigure}
	\caption{The mean estimation error of recovering the target image~$F$, as a function of the SNR, by approximate EM (Algorithm~\ref{alg:approx_EM}) and autocorrelation analysis. Evidently, the approximate EM algorithm outperforms autocorrelation analysis.}
\label{fig:noise_experiment}
\end{figure}

\section{Numerical experiments}
\label{sec:numeric}
\begin{figure}[!tb]
	\begin{subfigure}[ht]{.49\columnwidth}
		\centering
		\includegraphics[width=1\columnwidth]{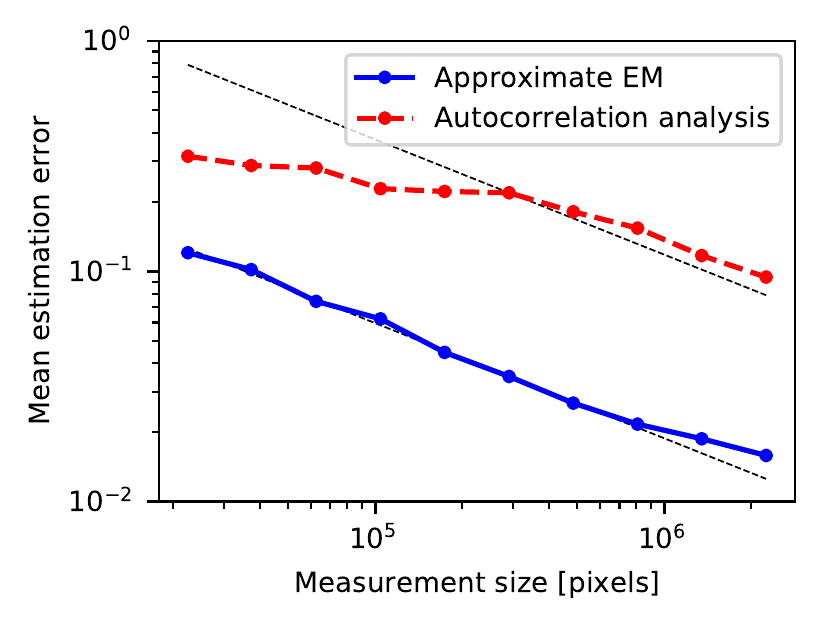}
	\end{subfigure}
	\begin{subfigure}[ht]{.49\columnwidth}
		\centering
		\includegraphics[width=1\columnwidth]{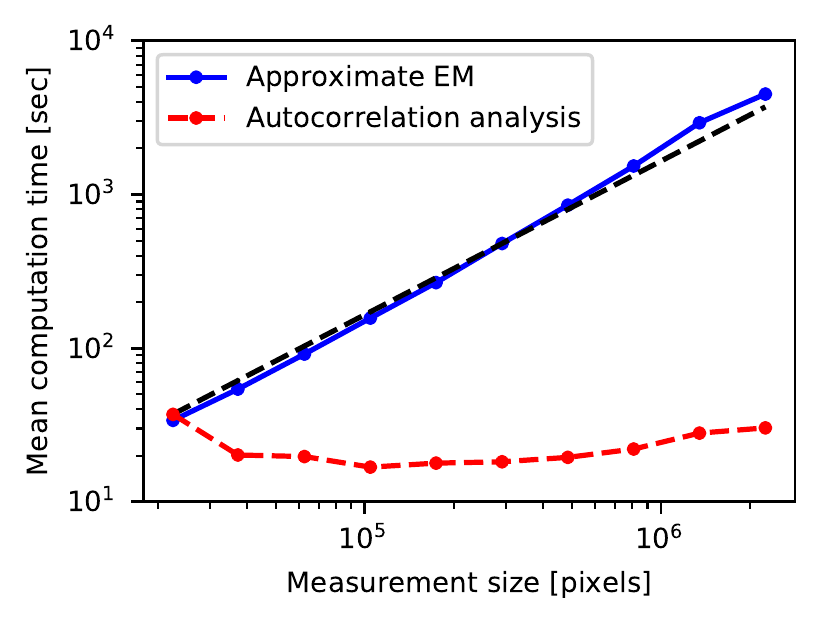}
	\end{subfigure}
	\caption{The mean estimation error of recovering the target image~$F$~(left) and running time (right), as a function of the measurement size~$N^2$ by approximate EM and autocorrelation analysis. For the estimation error, the black dashed lines illustrate a slope of~$-1/2$, as predicted by the law of large numbers. For the computation time, it illustrates a slope of~$1$, implying a linear increase in computation time, since the number of patches~\mbox{$N_d = N^2 / L^2$} grows linearly in~$N^2$.}
\label{fig:size_experiment}
\end{figure}

\begin{figure}[!tb]
	\begin{subfigure}[ht]{.49\columnwidth}
		\centering
		\includegraphics[width=1\columnwidth]{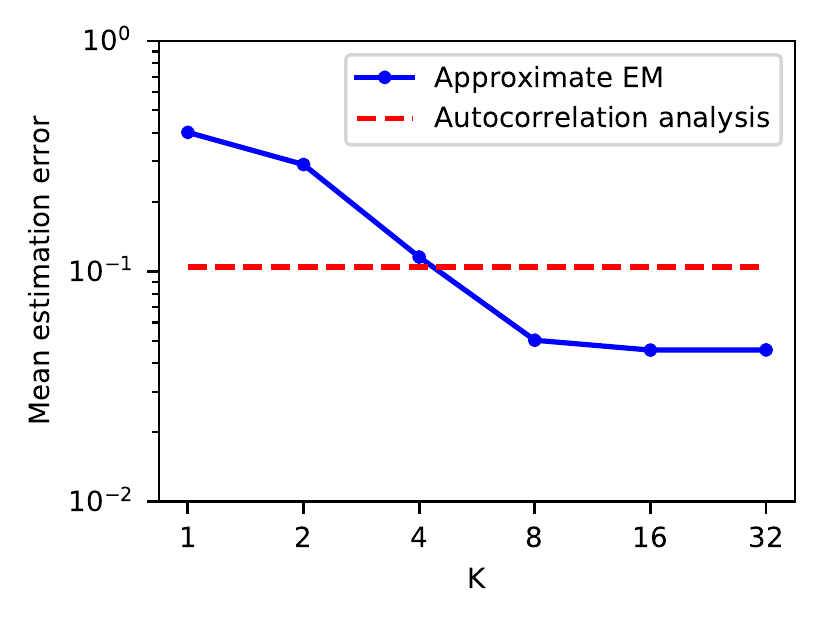}
	\end{subfigure}
	\hfill
	\begin{subfigure}[ht]{.49\columnwidth}
		\centering
		\includegraphics[width=1\columnwidth]{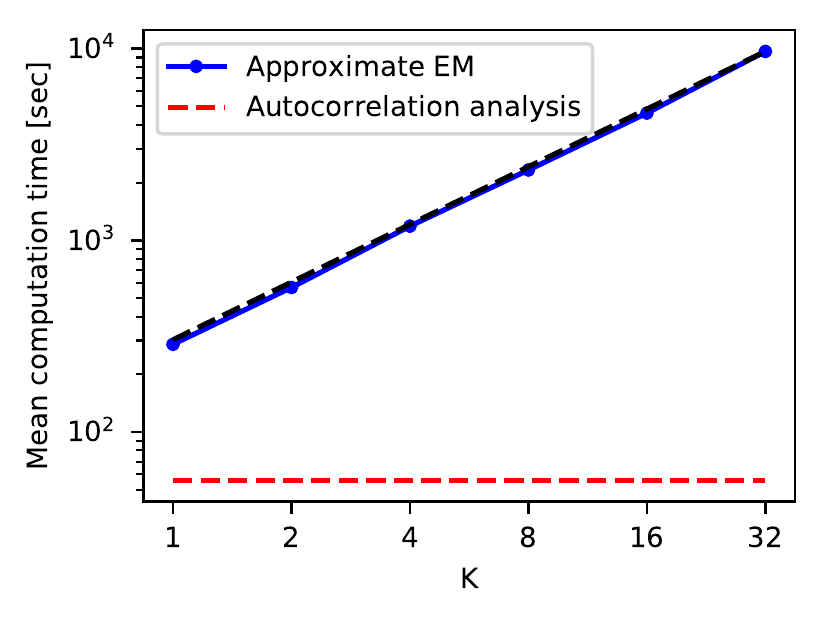}
	\end{subfigure}
	\caption{The mean estimation error of recovering the target image~$F$ (left) and corresponding running time (right), as a function of the size of the search space of rotations~$K$, by approximate EM. The recovery error and running time using autocorrelation analysis are marked by  dashed  red lines. For the running time, the black dashed line illustrates a slope of~$1$, which implies a linear increase in computation time, as the number of computations per patch depends linearly on~$K$.}
	\label{fig:discretization_experiment}
\end{figure}

In this section, we present numerical results for the approximate EM described in Section~\ref{sec:EM}. As a baseline, we compare the results against autocorrelation analysis with the first three autocorrelations based on the framework (and code) of~\cite{marshall2020image, bendory2021multi, kreymer2021two}. To take the in-plane rotation symmetry into account, we measure the estimation error by~$\min_{\phi \in [0, 2\pi)} \frac{\|\alpha^* - \alpha_{\phi}\|_2}{\|\alpha^*\|_2},$ where~$\alpha^*$ is the true vector of expansion coefficients, and~$\alpha_{\phi}$ is the vector of coefficients of the estimated image, rotated by~$\phi$. In all experiments, the measurements were generated according to~(\ref{eq:model}) with density~\mbox{\rev{$\gamma := p \frac{\pi n^2}{N^2} = 0.04$}}. The rotations  were drawn from a uniform distribution on~$[0, 2\pi)$, while the search space of the EM was discretized with a parameter $K$ according to~\eqref{eq:Phi_set}. The target images are of diameter~\mbox{$L = 5 \text{ pixels}$}. Each entry of the target images was drawn i.i.d.\ from a uniform distribution on~$[0,1]$. Then, each image was normalized such that~\mbox{$\|F\|_\text{F} = 10$}, and expanded using its first~$10$ coefficients as in~\eqref{eq:expansion}. The initializations of the EM and autocorrelation analysis iterations were drawn from the same distribution as the ground truth images, and~\mbox{$\gamma_{\text{init}} = 0.03$}. If the algorithms were initialized from several random  points, we calculated the  error of the image estimate whose likelihood is maximal (for approximate EM), or  whose objective function is minimal (for autocorrelation analysis). Figures~\ref{fig:noise_experiment}, \ref{fig:size_experiment} and~\ref{fig:discretization_experiment} present the mean error over~$40$ trials. The code to reproduce all experiments is publicly available at~\url{https://github.com/krshay/MTD-2D-EM}.

\vspace{-5pt}
\subsection{Recovery error as a function of the SNR}
\label{subsec:exp_SNR}
Figure~\ref{fig:noise_experiment} presents recovery error as a function of the SNR. The measurements are of size~\mbox{$N^2 = 2500^2 \text{ pixels}$}, and we use~\mbox{$K = 8$} possible rotations~\eqref{eq:Phi_set}, and~$5$ random initial guesses for~$\alpha$. To save computation time (see Section~\ref{sec:conclusion}), we initialized the approximate EM algorithm using the estimate achieved by autocorrelation analysis. We achieve a significant improvement in recovery accuracy using approximate EM, even though the search space of rotations is coarsely sampled.

In an additional numerical experiment with~$\text{SNR} = 2$ (low SNR, see Figure~\ref{fig:Micrographs_noise}), $N^2 = 10000^2 \text{ pixels}$, and~\mbox{$K = 16$}, the relative error of approximate EM was~$\text{0.017}$, whereas the error of autocorrelation analysis was~$0.073$.

\vspace{-5pt}
\subsection{Recovery error as a function of the measurement size}
\label{subsec:exp_size}

Figure~\ref{fig:size_experiment} presents recovery error and running time as a function of the measurement size $N^2$, with~\mbox{$\text{SNR} = 5$},~\mbox{$K = 16$}, and~$5$ random initial guesses for~$\alpha$. Using approximate EM, the error decays \mbox{as~$1 / \sqrt{N^2}$}. The same trend is visible also for autocorrelation analysis for sufficiently large measurements. We achieve a significant improvement in recovery accuracy using approximate EM. However, the computation time of our method is greater, and grows linearly with the measurement size. In particular, autocorrelation analysis is faster since it requires a single pass over the measurement, while the computational complexity of matching an image to the observed autocorrelations scale with~$L^2\ll N^2$.

\vspace{-5pt}
\subsection{Recovery error as a function of discretization of rotations}
\label{subsec:exp_discretization}
Figure~\ref{fig:discretization_experiment} presents recovery error and running time as a function of~$K$, the size of the search space of rotations, for measurements with~\mbox{$\text{SNR} = 5$}, $N^2 = 1500^2 \text{ pixels}$, and one  initial guess for~$\alpha$. Remarkably, even when the EM searches over only~$4$ rotations (recall that the rotations are drawn from a continuous distribution), the obtained estimation error is similar to the estimation error of autocorrelation analysis, which takes all possible (infinitely many) rotations into account. As expected, the computation time grows linearly with the parameter~$K$. This implies that one can save running time by coarsely sampling  the search space of rotations without severely degrading  the estimation quality.

\section{Conclusion}
\label{sec:conclusion}
This paper is motivated by the effort of reconstructing small~\mbox{3-D} molecular structures using \mbox{cryo-EM}, below the current detection limit~\cite{bendory2018toward}. The main contribution of this paper is in introducing an approximate EM scheme for the 2-D MTD problem, and comparing it numerically to autocorrelation analysis. The numerical experiments show an improvement in estimation accuracy, but at the cost of computational time. As Figure~\ref{fig:discretization_experiment} shows, the parameter~$K$ provides an accuracy-running time trade-off. A possible improvement is to increase the resolution of the search space as the iterations progress~\cite{lan2020multi}, or to design a branch-and-bound algorithm that accelerates EM by quickly and inexpensively ruling out large regions of the search space that  have very low probability to contain the optimum of the objective function~\cite{punjani2017cryosparc}.

Our ultimate goal is to develop an approximate EM scheme for recovering small molecular structures using \mbox{cryo-EM}~\cite{bendory2018toward}. In order to achieve a computationally efficient algorithm for the \mbox{3-D} case of \mbox{cryo-EM}, parallel processing and randomized algorithms, such as stochastic or online EM~\cite{nielsen2000stochastic, chen2018stochastic, liang2009online, cappe2009line, cappe2011online}, must be utilized. Further accelerations can be achieved by applying the EM algorithm on a lower dimensional representation of the data~\cite{dvornek2015subspaceem}, and initializing the  EM iterations with efficient computational techniques, such as autocorrelation analysis or stochastic gradient descent~\cite{punjani2017cryosparc}. Moreover, adding a prior on the target image is expected to improve robustness at the cost of  possible model bias. Another research direction is replacing EM by more intricate techniques that aim to approximate the posterior distribution, such as variational inference~\cite{blei2017variational} or variational auto-encoders~\cite{rosenbaum2021inferring}.


\end{document}